# Improving nuclear data evaluations with predictive reaction theory and indirect measurements

*Jutta* Escher[1]*, *Kirana* Bergstrom[2], *Emanuel* Chimanski[3], *Oliver* Gorton[4], *Eun Jin* In[1], *Michael* Kruse[1], *Sophie* Péru[5], *Cole* Pruitt[1], *Rida* Rahman[6], *Emily* Shinkle[7], *Aaina* Thapa[1], *and Walid* Younes[1]

[1]Lawrence Livermore National Laboratory, Livermore, CA 94551, USA
[2]University of Colorado Denver, Denver, CO 80204, USA
[3]Brookhaven National Laboratory, Upton, NY 11973, USA
[4]San Diego State University, San Diego, 92182, USA
[5]CEA, DAM, DIF, F-91680 Arpajon, France
[6]University of Tennessee Knoxville, Knoxville, TN 37996, USA
[7]University of Illinois Urbana-Champaign, IL 61801, USA

**Abstract.** Nuclear reaction data required for astrophysics and applications is incomplete, as not all nuclear reactions can be measured or reliably predicted. Neutron-induced reactions involving unstable targets are particularly challenging, but often critical for simulations. In response to this need, indirect approaches, such as the surrogate reaction method, have been developed. Nuclear theory is key to extract reliable cross sections from such indirect measurements. We describe ongoing efforts to expand the theoretical capabilities that enable surrogate reaction measurements. We focus on microscopic predictions for charged-particle inelastic scattering, uncertainty-quantified optical nucleon-nucleus models, and neural-network enhanced parameter inference.

## 1 Addressing data needs for unstable isotopes with indirect measurements

Neutron-induced reactions play an important role for nuclear astrophysics and applications. Many required reaction cross sections are unknown and extremely difficult to determine experimentally, as their measurement involves colliding neutrons with short-lived or highly-radioactive targets. For compound-nuclear reactions, Hauser-Feshbach (HF) statistical reaction calculations are used to determine the cross sections of interest. However, HF calculations require a number of nuclear physics inputs and suffer from large uncertainties in the absence of data that provide constraints on the models and parameters used. Reactions that are difficult to measure, but are of interest, include neutron capture, inelastic neutron scattering, (n,2n), and (n,f) reactions, as well as reactions with charged particles in the exit channel.

When direct measurements are not possible, indirect approaches such as the surrogate reaction method [1] may be used. This approach was originally introduced to infer (n,f) cross sections from inelastic scattering or transfer reactions, but significant theory developments have made it possible to also determine (n,γ) cross sections [2-4].

The surrogate reaction method is designed to provide constraints for the models describing the decay of the compound nucleus B* through which the reaction (a + A → B* → c + C) proceeds. When the entrance channel of interest (a + A) cannot be produced experimentally, one can use a surrogate reaction (d + D → B* + b) to produce the same compound nucleus B* and observe the decay of B* into the channel of interest (c + C). For most compound reactions, the decay of B* depends on the spins and parities (Jπ) populated in the surrogate reaction, therefore theory is required to infer decay model constraints from the indirect measurement. This requires a description of the surrogate reaction and a calculation of the Jπ distribution $F(E_{ex},J,\pi)$ of B*, at relevant excitation energies $E_{ex}$, prior to decay. The $F(E_{ex},J,\pi)$ are used to predict the observables from the indirect measurement. By adjusting the decay model and associated parameters to reproduce the data, the method yields the desired constraints. Subsequently, the cross section for the desired reaction can be calculated.

To make the surrogate reaction method more broadly applicable and to increase the impact of the resulting cross sections, further advances in nuclear reaction theory and a focus on parameter inference are required. Here, we will highlight recent developments that address both aspects.

## 2 Charged-particle inelastic scattering as surrogate reaction mechanism

Recent applications of the surrogate reaction method to obtain neutron capture cross sections demonstrated that modelling the (p,d) and (d,p) transfer reactions utilized involves knowledge of nuclear structure effects and higher-order reaction contributions, such as two-step transfers or breakup-fusion [2,3,5]. Inelastic scattering opens opportunities to reach additional compound nuclei as well as excitation energies high enough (tens of MeV) to study decays relevant to (n,n') and (n,2n) reactions. It is therefore important to build on and expand the initial modelling for charged-particle scattering [4,6-8] to enable a broader range of experiments.

---

* Corresponding author: escher1@llnl.gov

To model the direct inelastic scattering process relevant to such a surrogate reaction measurement, we utilize an integrated nuclear structure and reaction framework. To describe the properties of the ground and various excited states involved in the surrogate reaction, we utilize a structure theory that is able to predict both low-lying and highly-excited states and that can be used for medium-mass and heavy nuclei across the isotopic chart: Hartree-Fock-Bogoliubov (HFB) and quasi-particle Random Phase Approximation (QRPA) calculations provide us with ground-state properties and transition densities for states reached by inelastic scattering. Our particular implementation is built on the Gogny interaction [9,10] and utilizes an axially-symmetric deformed basis [11,12]. We use a folding approach and convolute the transition densities with an effective projectile-target nucleon interaction to obtain transition potentials suitable for use in a coupled-channels code, such as FRESCO [13].

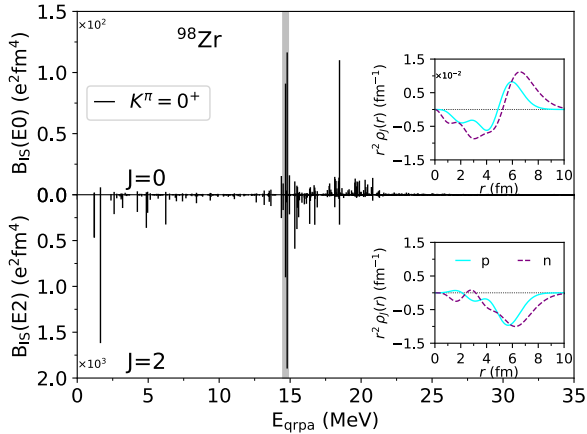

**Fig. 1.** Electromagnetic transitions of monopole and quadrupole character in the $K^\pi = 0^+$ QRPA component of deformed nucleus $^{98}$Zr. The inserts show radial transition densities for highlighted states near $E_{ex}$ = 15 MeV.

To test the quality of the structure predictions, we calculate properties for chains of isotopes, e.g. for stable Zr and Mo isotopes. Ground-state properties, including deformations and charge radii agree well with measured values (not shown). In Figure 1, we show a representative example of monopole and quadrupole excitations in the deformed nucleus $^{98}$Zr. The main part of the figure shows reduced transition probabilities for iso-scalar E0 (upper portion) and E2 (lower portion) excitations. Only $K^\pi=0^+$ QRPA components are shown, to demonstrate that there exists coupling between the E0 and E2 excitations for many states in this deformed nucleus. $K^\pi = \pm 1^+, \pm 2^+$ components (not shown) also contribute to E2 excitations, and are shifted in energy due to deformation effects. Radial transition densities for representative states are shown in the inserts. These radial functions are used to generate the transition potentials, which in turn are used in distorted-wave Born Approximation (DWBA) or coupled-channels (CC) direct-reaction calculations.

In addition to testing the structure model against experimental data, it is important to carefully consider the reaction mechanisms involved in predicting scattering observables. It has long been known from studies of giant resonances that inelastic scattering calculations of this type are only able to reproduce a fraction of the observed cross section [14]. The general approach in the giant-resonance studies is to introduce an *ad hoc* procedure for removing a large portion of the cross section from consideration and to focus on the portion deemed relevant for investigating the resonances [15]. This procedure is not adequate for surrogate reaction applications, as the unexplained part of the cross section is expected to contribute to CN formation and, importantly, to the spin-parity distribution that is needed for a proper implementation of the method. A better understanding of CN formation in inelastic scattering is required to address this issue.

## 3 Parameter inference and uncertainty quantification for cross sections from indirect data

Applications require not only reaction cross sections, but also estimates of the associated uncertainties. Markov-Chain Monte-Carlo (MCMC) methods allow for the propagation of uncertainties and determination of correlations. Neural networks are being explored to speed up the parameter inference process. While the focus of the surrogate reaction approach is on obtaining constraints for the decay of the CN, the optical model potentials used, and the associated uncertainties deserve additional consideration.

### 3.1 Parameter inference using Hauser-Feshbach theory and Markov-Chain Monte-Carlo

The objective of the surrogate reaction approach is to determine constraints for Hauser-Feshbach reaction calculations. More specifically, we solve the inverse problem of propagating uncertainty for dependent variables (HF model of surrogate reaction data) into uncertainty for the independent variables (decay parameters). In earlier applications a Bayesian approach with a simple Monte-Carlo parameter variation was employed [2,3]. An improved algorithm, shown schematically in Figure 2, utilizes a MCMC method to obtain posterior distributions for parameters and their correlations. These distributions can then be sampled to calculate the desired cross section, along with covariances.

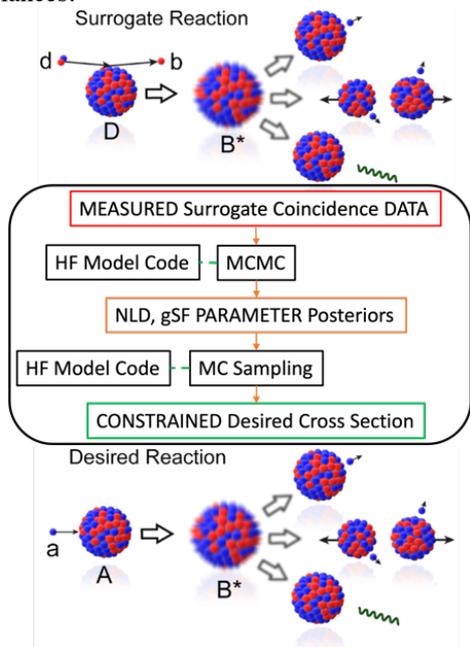

**Fig. 2.** Parameter inference for Hauser-Feshbach decay models using a Markov-Chain Monte-Carlo approach. MCMC is used to solve the inverse problem of obtaining parameters to describe the data. MC can then be used in the forward propagation of newly determined constraints on model parameters into predictions for the desired cross section.

### 3.2 Speeding up parameter inference and interpretation with neural networks

Typical Hauser-Feshbach calculations are not computationally challenging. While pre-calculating transmission coefficients with coupled-channels codes may take a little longer, many HF calculations run in seconds or minutes. An exception are runs with HF codes that implement a Monte-Carlo computational scheme, rather than a deterministic approach [16,17]. Parameter fitting with MC-based codes can take considerable time, depending on the number of parameters to be adjusted and the number of reactions under consideration. Surrogate reaction applications require both decay calculations to fit the decay parameters and calculations for the desired reaction cross sections. It therefore becomes important to consider efficient alternatives to full HF calculations.

We are exploring neural networks as potential emulators for Hauser-Feshbach decay and reaction calculations. In a first step, we use the HF code YAHFC [17] to generate training data for a neural network. The data consisted of 1848 YAHFC input/output pairs, each with 13 input parameters and 5 output curves of calculated surrogate coincidence probabilities with 155 data points. We trained a feed-forward network with one 60-node hidden layer, mapping the 13 input features to 155 outputs. The trained 'proxy' model is then used in conjunction with surrogate reaction data and the MCMC framework described above to determine a posterior parameter distribution. The top panel of Figure 3 shows a comparison of the calculated probabilities with the data. The curves give the distribution resulting from this process.

In the second step, we use the same HF code to generate training data for calculating the desired cross section, here $^{95}$Mo(n,$\gamma$). A second 'proxy' model is then trained and subsequently used to sample the posterior distribution from the first step and generate cross section predictions. Preliminary results are shown in the bottom panel of Figure 3.

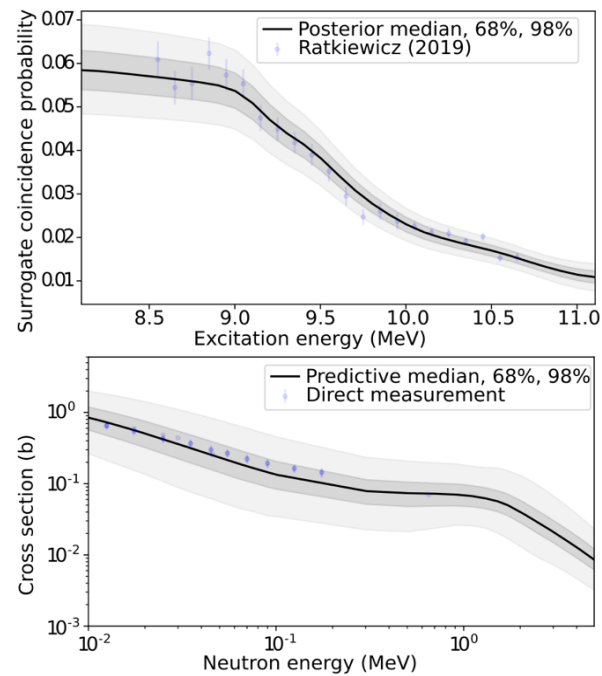

**Fig. 3.** Parameter inference for Hauser-Feshbach (HF) decay models using predictions from neural network emulators for HF calculations (Preliminary Results). Top panel: Posterior distribution of coincidence probabilities predicted by the first neural network HF emulator. Bottom panel: Posterior distribution of the cross section predicted by the second HF emulator. In both cases, the black curve gives the median, the dark and light grey bands indicate the central 68th and 98th percentiles of the distributions.

The initial results from the neural-network approach are encouraging: The trained models agree well with full HF calculations, for both the decay and the cross section calculations. The time to complete the training, parameter inference, and final predictions for the cross section, within the MCMC framework, is only about one sixtieth (1/60) of the time it takes for the equivalent process that uses the full HF calculations. More work is needed to explore cases involving reactions with additional decay channels and more parameters.

### 3.3 Uncertainty-quantified optical potentials

Present applications of the surrogate reactions approach do not incorporate uncertainties due to the optical model potentials (OMPs) used in the calculations. In the cases studied, all of which involve nuclei near the valley of stability, unconstrained level densities and gamma-ray strength functions are known to be the primary source of uncertainty. Moreover, most nuclear data evaluations, which are adjusted to *directly-measured* data, do not account for optical-model uncertainties either. To achieve a more complete assessment of the reliability of calculated cross sections, it is necessary to develop uncertainty-quantified optical-model potentials [18].

We have developed updated parameterizations and well-calibrated uncertainties [19] for two widely-used nucleon-nucleus OMPs, the Chapel-Hill CH89 OMP and the Koning-Delaroche KD OMP [20,21]. A large data corpus of scattering observables was compiled. A new framework was developed that includes outlier identification and assessment of unaccounted-for uncertainty and a MCMC approach was employed to update the parameters and obtain complete covariance information. Scattering predictions from the parameter ensembles generated in this context show improved performance against the original (training) and newly measured (test) data (not shown). Case studies, such as the one shown in Figure 4, demonstrate the utility of the ensembles for propagating uncertainties into reaction calculations, here for proton capture on $^{87}$Sr.

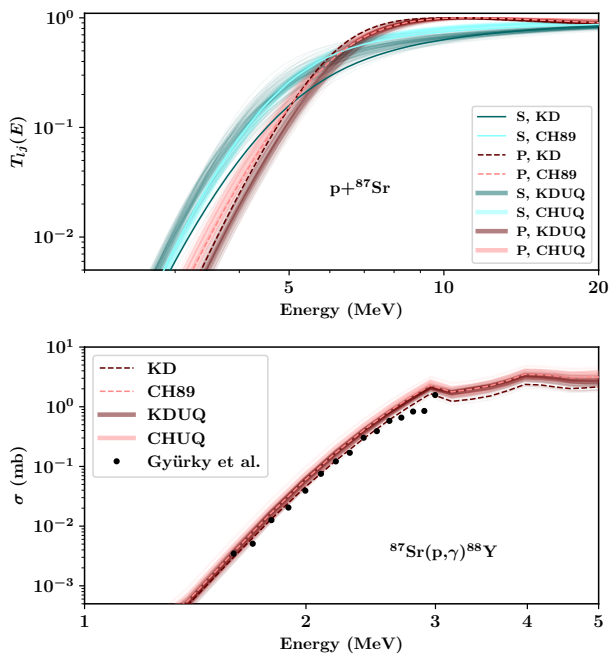

**Fig. 4.** Predictions for the p + $^{87}$Sr system, calculated after obtaining updated parameterizations CHUQ and KDUQ for the CH89 and KD potentials, respectively. Uncertainties have been propagated forward into essential nuclear data inputs, such as transmission coefficients (top panel) and capture cross sections (bottom panel). Figure adapted from Ref. [19].

## 4 Conclusions and outlook

The availability of reliable nuclear data with quantified uncertainties is essential for basic and applied science. The last decade has seen much progress in the development of theory tools that allow for more accurate evaluations for compound (statistical) nuclear reactions. Significant theory development aimed at describing surrogate reaction mechanisms have made it possible to determine neutron capture cross sections for short-lived targets indirectly. We have described ongoing developments to achieve predictions for charged-particle inelastic scattering, uncertainty-quantified optical nucleon-nucleus models, and neural-network enhanced parameter inference. Together with new experimental efforts, these advances will make this indirect approach more accurate and more broadly applicable to address nuclear data needs.

This work was performed under the auspices of the U.S. Department of Energy by Lawrence Livermore National Laboratory under Contract DE-AC52-07NA27344, with support from LDRD projects 19-ERD-017, 20-ERD-030, and 22-LW-029. We gratefully acknowledge support from the Defense Science and Technology internship (DSTI) and the Seaborg Institute for K. Bergstrom, O. Gorton, E. Shinkle, and R. Rahman.

## References

1. J.E. Escher, J.T. Harke, F.S. Dietrich *et al.*, Rev. Mod. Phys. **84**, 353 (2012).
2. J.E. Escher, J.T. Harke, R.O. Hughes, *et al.*, Phys. Rev. Lett. **121**, 052501 (2018).
3. A. Ratkiewicz, J.A. Cizewski, J.E. Escher, *et al.*, Phys. Rev. Lett. **122**, 052502 (2019).
4. R.P. Sánchez, B. Jurado, V. Méot, *et al.*, Phys. Rev. Lett. **125**, 122502 (2020).
5. G. Potel, G. Perdikakis, B.V. Carlson, *et al.*, Eur. Phys. J. A **53**, 178 (2017).
6. G.P.A. Nobre, F.S. Dietrich, J.E. Escher *et al,*, Phys. Rev. Lett 105, 202502 (2010)
7. G.P.A. Nobre, F.S. Dietrich, J.E. Escher *et al.*, Phys. Rev. C 84, 064609 (2011)
8. M. Dupuis, G. Haouat, J.-P. Delaroche *et al.*, Phys. Rev. C **100**, 044607 (2019).
9. J. Dechargé and D. Gogny, Phys. Rev. C **21**, 1568 (1980).
10. S. Goriely, S. Hilaire, M. Girod, and S. Péru, Phys. Rev. Lett. **102**, 242501 (2009).
11. S. Péru and M. Martini, Eur. Phys. J. A **50**, 88 (2014).
12. W. Younes, D. Gogny, and J.F. Berger, *A Microscopic Theory of Fission Dynamics Based on the Generator Coordinate Method* (Springer, Cham, 2019).
13. I.J. Thompson, Comput. Phys. Rep. **7**, 167 (1988).
14. S.F. Tsai and G.F. Bertsch, Phys. Rev. C **11**, 1634 (1975).
15. F.E. Bertrand, Nucl. Phys. A **354**, 129 (1981).
16. T. Kawano, Springer Proceedings in Physics 254, 27 (2021).
17. W.E. Ormand and K. Kravvaris, LLNL Report, LLNL-TR-821653 (2021).
18. C. Hebborn, F.M. Nunes, G. Potel, *et al.*, J. Phys. G (in print, 2023), arXiv:2210.07293v2 (2023).
19. C. Pruitt, J.E. Escher, and R. Rahman, Phys. Rev. C **107**, 014602 (2023).
20. R.L. Varner, W.J. Thompson, T.L. McAbee, *et al.*, Phys. Rep. **201**, 57 (1991).
21. A.J. Koning and J.-P. Delaroche, Nucl. Phys. A **713**, 231 (2003).